%% file: rule2009.tex
\documentclass{eptcs}

\providecommand{\event}{SOS 2007} 
\providecommand{\volume}{??}
\providecommand{\anno}{20??}
\providecommand{\firstpage}{1}
\providecommand{\eid}{??}\usepackage{breakurl}

\usepackage[utf8]{inputenc}
\usepackage{amsfonts,amsmath,amssymb,amsthm}
\usepackage{graphicx,color,url,listings,multicol}
\newtheorem{definition}{Definition}
\newtheorem{example}{Example}

\def\titlerunning{Modeling and Reasoning over Distributed Systems using AOGGs}
\def\authorrunning{Machado R., Heckel R., Ribeiro L.}
\title{Modeling and Reasoning over Distributed Systems using Aspect-Oriented Graph Grammars}
\author{Rodrigo Machado
\institute{Univ. Federal do Rio Grande do Sul\\Porto Alegre, Brazil}
\email{rma@inf.ufrgs.br}
\and
Reiko Heckel 
\institute{Univ. of Leicester\\ Leicester, UK}
\email{reiko@mcs.le.ac.uk}
\and
Leila Ribeiro
\institute{Univ. Federal do Rio Grande do Sul\\Porto Alegre, Brazil}
\email{leila@inf.ufrgs.br}
}

\input{defs.tex}

\begin{document}

\maketitle

\begin{abstract}
Aspect-orientation is a relatively new paradigm that introduces abstractions to modularize the implementation of system-wide policies. It is based on a composition operation, called aspect weaving, that implicitly modifies a base system by performing related changes within the system modules. Aspect-oriented graph grammars (AOGG) extend the classic graph grammar formalism by  defining aspects as sets of rule-based modifications over a base graph grammar. Despite the advantages of aspect-oriented concepts regarding modularity, the implicit nature of the aspect weaving operation may also introduce issues when reasoning about the system behavior. Since in AOGGs aspect weaving is characterized by means of rule-based rewriting, we can overcome these problems by using known analysis techniques from the graph transformation literature to study aspect composition. In this paper, we present a case study of a distributed client-server system with global policies, modeled as an aspect-oriented graph grammar, and discuss how to use the AGG tool to identify potential conflicts in aspect weaving.
\end{abstract}

\section{Introduction}

Aspect-oriented programming \cite{Kiczales1997} is a relatively new paradigm which introduces abstractions to modularize the implementation of system-wide policies. It is based on a composition operation, called aspect weaving, that modifies a base system globally according to structural rules such as, for instance, ``register in a global log all modifications in the value of the field \emph{x} of class \emph{C}''. As characterized in \cite{Filman2000}, an aspect is a module that \textit{i)} identifies in other modules sets of execution points, which are called \emph{pointcuts}, and \textit{ii)}  define transformation rules associated with pointcuts. Those rules are called \emph{advices}.
Given a pointcut language expressive enough, the implementation of global policies becomes modular and consistent as the system evolves, i.e. new modules will abide by the global policy in the same way as the current ones. 

Those advantages stimulated the adoption of aspect-oriented programming in software development. Several
languages now have aspect-oriented extensions, the most popular being the Java superset AspectJ \cite{Kiczales2001}. Moreover, the usage of AOP-related 
concepts also started to appear in languages for system modeling, such as UML diagrams \cite{Junior2009}. However, the wide usage of aspect-oriented concepts also introduces issues in the software development process. The implicit modifications caused by aspect weaving may result in behaviors which are difficult to identify by source code analysis. Also, when
the system has more than one aspect they may interfere with each other, resulting in different final systems according
to the order they are combined. To deal with those problems, the developer needs proper models 
to reason consistently about the aspect influence. On the formal side, several aspect-oriented calculi have been proposed to characterize aspect interference over programming languages \cite{Walker2003, Jagadeesan2006, Djoko2006, Clifton2006}. On the implementation side, integrated development environments offer support to new views related to aspect weaving \cite{AJDT}. However, outside the scope of source-code level aspects, there are still few models and techniques available to reason about aspect-oriented diagrams.

The current proposals for studying aspect weaving over diagrams have a strong connection with graph transformation systems and graph grammars, models where the system state is represented by a graph, and its execution, by the application of graph rewriting rules. 
Given the natural representation of diagrams as graphs, both aspect-oriented systems and graph transformation systems share some common characteristics: pointcuts resemble matches for graph rules, and advices resemble graph rules themselves. In \cite{Machado2009}, aspect-oriented graph grammars (AOGGs) were proposed as an extension of the traditional graph grammars, where aspects were modeled as second-order transformations over the original specification. The advantage of this approach is that the same rewriting mechanism is used for both aspect weaving and base system execution, allowing to relate them formally. However, up to now it was not shown how to reason about AOGG models. In this work, we propose the use of AGG \cite{AGG}, an attributed graph grammar specification and analysis tool to reason about AOGG models. Since AGG does not support directly AOGGs, we propose an encoding procedure to convert graph grammar specifications into a single typed graph, with aspects being modeled as sets of graph rewriting rules. Then, we use the analysis functionalities of AGG to study conflicts between advices and aspects in the encoded AOGG system.

The text is organized as follows: in Section~\ref{sec:gg}, we review  the graph grammar model, and 
introduce the base client-server example. In Section~\ref{sec:aogg}, we recall aspect-oriented graph grammars and present 
examples of aspects over the base system. In Section~\ref{sec:encoding} we present the encoding of AOGGs as typed graphs
in order to use the AGG tool. We discuss the use of critical pair analysis to study the
aspect weaving in the encoded system in Section \ref{sec:analyze}.  Final remarks, related work and future steps are discussed in Section~\ref{sec:conclusion}. 

\section{Graph Grammars}
\label{sec:gg}

A typed graph grammar is a visual model where the system state is represented by a graph and the system behavior is described
by means of rule-based graph rewriting. Formally, a typed graph grammar is a tuple $\<T, G_0, P, \pi \>$, 
where \emph{i)} $T$ is a type graph defining the allowed kinds of nodes and edges within the specification;
\emph{ii)} the $T$-typed graph $G_0$ is the initial state of the system; \emph{iii)} $P$ is a set of rule names, and \emph{iv)} the function $\pi : P \to Rules(T)$ maps every rule name to is respective $T$-typed graph transformation rule. 
In this work we follow the double-pushout (DPO) approach to graph transformation \cite{Ehrig1973}, where a graph transformation rule is represented as a monic span (a pair of injective morphisms with the same source) $L \stackrel{l}\leftarrowtail K \stackrel{r}\rightarrowtail R$ in the category \Typed{T}\  ($T$-typed graphs and type-preserving graph homomorphisms).
The left-hand side (LHS) graph $L$ represents the pattern to be found in order to apply the rule. The right-hand side (RHS) graph $R$ defines new graph elements to be added by the rule. The interface graph $K$ and graph inclusions $l$ and $r$ identify elements in $L$ and $R$.  Elements in $L$ but not in $l(K)$ are said to be \emph{deleted}, elements
in $R$ but not in $r(K)$ are said to be \emph{created}, and elements in $K$ are \emph{preserved} or \emph{read}.

\begin{figure}[htbp]
\centering
\includegraphics[scale=0.8]{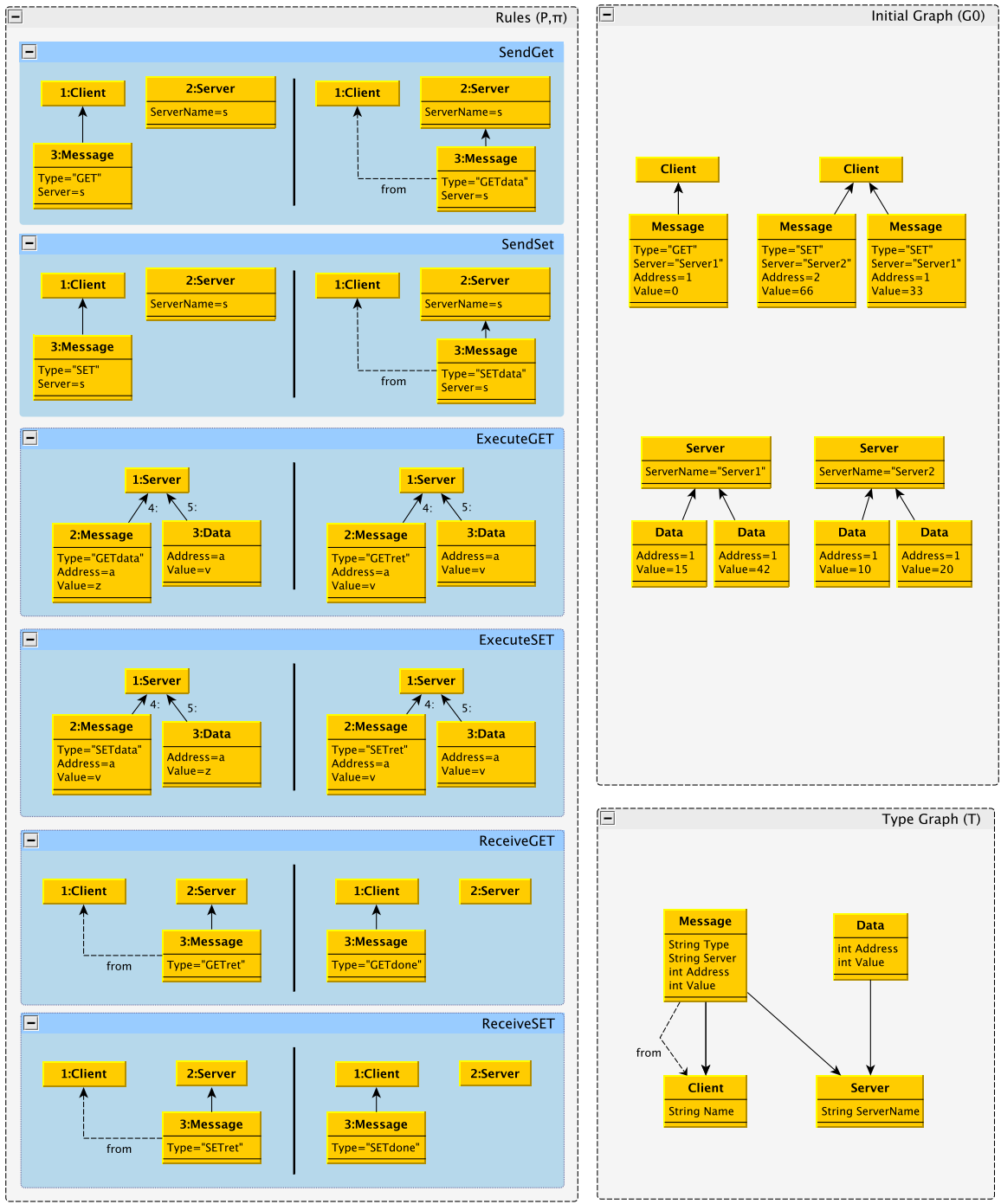}
\caption{Client-server system as a graph grammar.}
\label{fig:gragra}
\end{figure}

\begin{example}[{Graph grammar}]
\label{ex:base}
Figure~\ref{fig:gragra} depicts a distributed client-server system modeled as a graph grammar.
The type graph \texttt{T} defines four kinds of nodes (\texttt{Client}, \texttt{Server}, \texttt{Data} and \texttt{Message}) and  four kinds of edges.
The initial graph $G_0$ defines the initial state of the system: two clients with messages to be sent 
and two data servers. Clients may retrieve  values from servers or update stored data by means of message exchange.
The system rules model this interaction as follows: a \texttt{GET} message is sent to the server by \texttt{SendGET}, then the data is retrieved by \texttt{ExecuteGET}, and it is returned to the client by \texttt{ReceiveGET}.
The rules \texttt{SendSET}, \texttt{ExecuteSET} and \texttt{ReceiveSET} work in a similar way for messages
that update data elements. Regarding the notation, rules are presented only by their left-hand side
and right-hand side graphs. Common elements in $K$ are simply marked to be the same in $L$ and $R$
using numbered prefixes ($1\!\!:$, $2\!\!:$, \ldots). It is also important to notice that
this example uses attributed graph nodes, which is an extension of the basic typed graph grammar formalism.
\end{example}

In order to apply a graph transformation rule $L \gets K \to R$ over a graph $G$ it is required that 
there is some \emph{match} in $G$ for elements of $L$. Matches are defined as graph homomorphism $m: L \to G$.
Then, a rule application (or direct derivation) from graph $G$ to graph $H$ using rule $r$ and 
match $m$ (denoted $G \stackrel{r,m}{\Longrightarrow} H$) is defined as the following diagram in the 
category \Typed{T}, where both squares are pushouts. 
\begin{center}
\(
\xymatrix@=1.2cm{ 
      *+<1em>{L}  \ar @{}[dr]|{(1)} \ar@{>->}[d]_{m}
    & *+<1em>{K}  \ar@{>->}[l]_{l}  \ar@{>->}[r]^{r}  \ar[d]_{k} \ar@{}[dr]|{(2)}
    & *+<1em>{R} \ar[d]^{m^*}
\\    *+<1em>{G}
    & *+<1em>{D} \ar[l]^{l^*} \ar[r]_{r^*}
    & *+<1em>{H} \\
}\)
\end{center}
In the DPO approach, rule application relies on finding $G \stackrel{l^*}\gets D \stackrel{k}\gets K$ such 
that the square to the left is a pushout. This may not always be possible depending on the rule
and match. For instance, when the rule tries to delete a node that has incident edges not in $m(L)$, 
or when the match identifies preserved elements with deleted ones. 

The sequential behavior of a graph grammar is
given by the successive application of graph rules starting from
the initial graph. Usually, both rules and matches are nondeterministically chosen at each step. 
Operationally, this works as follows: 

\begin{enumerate}
\item Set $G_0$ as the current graph. 
\item Find in the current graph all possible matches for rules in the set $P$ of rule names.
\item Verify application conditions for matches, defining the set of valid pairs (rule,match).
\item If there is no valid pair at all, then STOP. Otherwise, non-deterministically choose a pair (rule,match) to be applied.
\item Delete from $G$ all matched elements in $L \setminus K$. This will generate a graph $D$.
\item Create in $D$ all elements in $R \setminus K$. This will generate a graph $H$.
\item Set $H$ as the current graph. Return to step 2.
\end{enumerate}

Graph grammars are particularly useful to represent distributed systems since graph rules 
represent local transformations. This means that two or more rules may be applied in parallel
if they do not conflict with each other. Moreover, the rules are usually intuitive and
the control flow is data-driven, being guided by the graph topology. 

The AGG tool \cite{AGG} allows to create, run and analyze graph grammars. Moreover, it supports several
extensions to the basic typed graph grammar language, such as attributed graph elements, 
control flow mechanisms (rule layers, priorities), definition of positive and negative application conditions
for rules, among others. The AGG graph transformation engine follows the single-pushout (SPO)  approach  \cite{Lowe1993}, 
but it may also be configured to consider the conditions that characterize DPO transformations. 

\section{Aspect-Oriented Graph Grammars}
\label{sec:aogg}

Aspect-Oriented Graph Grammars (AOGGs) \cite{Machado2009} are graph grammars with aspects, i.e. modular descriptions of system-wide policies. Formally, an AOGG is a pair $\< \mc{G}, \Delta \>$, where $\mc{G} = \< T, G_0, P, \pi \>$ is a base graph grammar, and $\Delta = [A_1, A_2, \ldots, A_n ]$ is a sequence of graph aspects over $\mc{G}$. A graph aspect represents a set of modifications over the base graph grammar, and it is defined by a triple $\< D, t, g\>$, where $D$ is a set of graph advices, $t: T \hookrightarrow T'$ is an extension of the original type graph and $g: G_0 \hookrightarrow G'$ is an extension of the original initial graph. 

Graph advices are rules that modify graph transformation rules. Following the intuition provided by the DPO approach, graph advices are defined as monic spans $p \leftarrowtail i \rightarrowtail e$. The difference is that $p,i,e$ are themselves graph rules, and $l,r$ are rule morphisms. Rule morphisms are defined as triples of morphisms connecting the graph components of two rules, respecting the commutativity of the inner squares. This way, a graph advice correspond to the following commutative diagram in $T$-\Graph.
\[
\xymatrix@C=0.4cm@R=0.1cm@*+<2ex>{
& & R_p &  & & \ar@{>->}[lll] \ar@{>->}[rrr]R_i &  & & R_e \\
& \ar@{>->}[ur] \ar@{>->}[dl]  K_p &  & & \ar@{>->}[ur] \ar@{>->}[dl] \ar@{>->}[rrr] \ar@{>->}[lll] K_i & &  & \ar@{>->}[ur] \ar@{>->}[dl] K_e & \\
L_p &  & & \ar@{>->}[lll] \ar@{>->}[rrr] L_i & &  & L_e & & 
}
\]

To emphasize the distinction between advices and base system rules, the components of a graph advice receive specific names: \emph{pointcut} $p$, \emph{advice interface} $i$ and \emph{effect} $e$.  

\begin{example}[Log Aspect]
Suppose we want to implement a logging mechanism over the client-server system such that 
every operation leaves an execution trace over a global object (the system logger).
To implement this functionality, we should extend the type graph to introduce the logger type, 
initialize it in the initial graph and modify all rules to register their execution in this
global object. Using AOGG, all these modifications can be enclosed within one single aspect,
as shown in Figure~\ref{fig:aogg-log}. This aspect has only one advice, which has an empty pointcut.
The advice adds a \texttt{Logger} object to both the LHS and RHS of the rule such
that the occurrence on the RHS has the rule name appended to the \texttt{story} string.
Notice that this value update depends on the existence of a reflective operation \texttt{rulename()} 
which provides the name of the rule being executed. The empty pointcut matches all possible 
rules of the specification, thus after aspect weaving all of them will read and update the global 
log object.

\begin{figure}[htbp]
\centering
\includegraphics[scale=0.8]{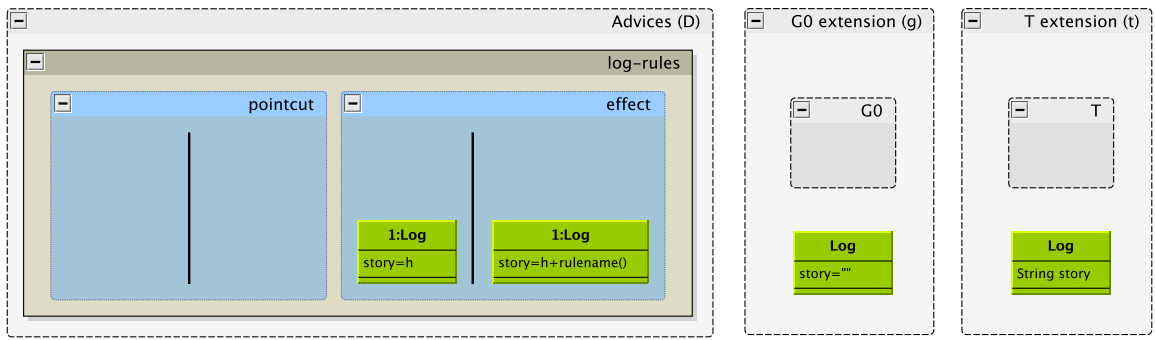}
\caption{Log aspect}
\label{fig:aogg-log}
\end{figure}
\end{example}

\begin{figure}[htbp]
\centering
\includegraphics[scale=0.8]{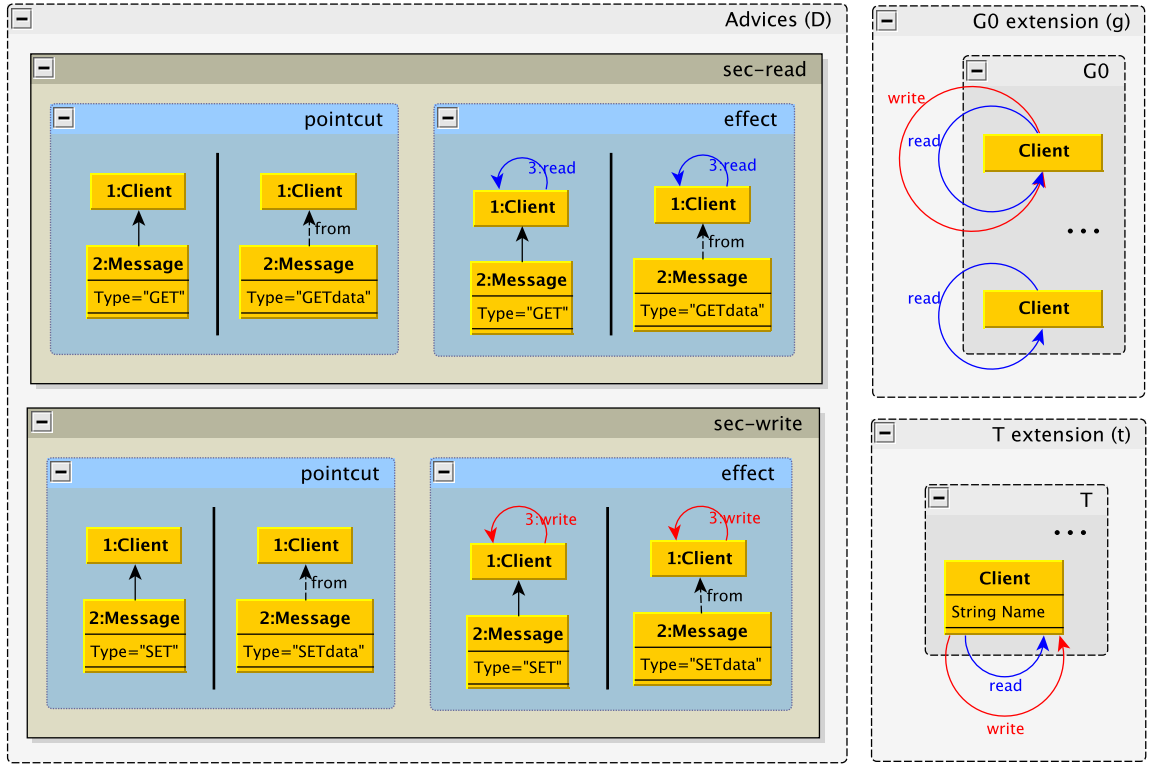}
\caption{Security aspect}
\label{fig:aogg-sec}
\end{figure}

\begin{example}[Security Aspect]
Another kind of system-wide modification is the introduction of access control for servers.
A simple implementation is to define two kinds of permissions
for clients: \texttt{Read} and \texttt{Write}. Those permissions, 
represented as endoarcs in client nodes,  affect
message exchanges. A \texttt{"SET"} message
may only be sent by a user with write privileges, 
while a \texttt{"GET"} message, only by a user with read privileges.  
The aspect depicted in Figure~\ref{fig:aogg-sec} implements this policy over the
base graph grammar of Example~\ref{ex:base}. 
The act of sending a message of a given kind (\texttt{SET} or \texttt{GET}) 
defines the pattern used in the \emph{pointcuts} of both advices. The effect is to augment the rules such that they
require the respective user permissions to be read in order to execute. In both advices the interface $i$ is
the same as the pointcut $p$.
\end{example}

As with graph rewriting, rule rewriting (advice application) is also defined as a double-pushout diagram.
The difference is that it uses the category \STyped{T}\ of $T$-typed rules and rule morphisms \cite{Machado2009}. 
Given a graph grammar $\mc{G} = \< T, G_0, P, \pi\> $ and a graph aspect $\<D, t, g\>$ over $\mc{G}$, with $t:T\to T'$ and $g: G_0 \to G_0'$, the result of weaving $A$ and $\mc{G}$ is the graph grammar $\mc{G}' = \< T', G_0', P', \pi'\>$. 
The rule set $(P',\pi')$ is calculated based on $(P,\pi)$ as the \emph{least} set of rules such that 
\be
  \tm If a rule in $(P,\pi)$ is not matched by any advice in $A$, them the rule is kept unchanged in  $(P',\pi')$.
  \tm If a rule is matched by at least one advice in $A$, all the results of one-step rewritings (considering all possible advices and matchings) become elements of $(P',\pi')$. In this case, the original rule does not appear in $(P',\pi')$.
\ee
This definition for aspect weaving makes advice rewriting non-reentrant, i.e. a given rule may not be modified
more than once per advice and match. This assures termination for the weaving process, even for advices that do not delete elements from rules. Finally, the semantics of an AOGG $\< \mc{G}, \Delta \>$ is given by its 
respective weaved graph grammar $\mc{G}^\Delta_W$, defined as the result of weaving all aspects of $\Delta$ over $\mc{G}$ in their respective order.

\section{Encoding AOGGs in AGG}
\label{sec:encoding}

The use of graph rewriting as an aspect weaving mechanism opens the possibility of using known
techniques from the graph transformation area to reason about aspects.
Those techniques allow the system modeler to identify potentially harmful behaviors
of the final weaved system, such as unintended aspect interactions, at early stages of the system development. 
Furthermore, existing tools such as AGG can be used to analyze aspect-oriented models.

Our approach consists in encoding the whole structure of a graph grammar specification as a single typed graph. 
For this, a representation of monic spans using graph elements is needed to encode individual rules.
Moreover, one needs to encode both initial graph and rule set while keeping their as distinguished elements of the specification. For this, we employ the traditional graph typing mechanism: the type graph for the 
encoding of $\mc{G} = \< T, G_0, P, \pi\> $ has two disjoint components. The first is the original type graph $T$, used to type the initial graph $G_0$. The second is $R(T)$, used to type the encoded representation of the
rule set. We start by defining the operation $R$ that calculates the second component based on the original type graph $T$.

\begin{definition}[Type graph rule encoding]
The operation $R: \Graph \to \Graph$ is defined as the composition $R_3 \circ R_2 \circ R_1$, where:
\bi
\tm $R_1$ encodes edges as nodes. The edges of the resulting graph follow the $source$ and $target$ functions of the input graph.
\tm $R_2$ creates three copies of the input graph (to represent $L$, $K$ and $R$). Each node in $K$ is connected to its copy in $L$ and $R$ by means of new edges.
\tm $R_3$ creates a new node in the graph together with edges connecting all the original nodes to this node. This node represents the identity of the rule.
\ei
\end{definition}

\begin{example}[Type graph rule encoding]
\label{ex:r}
Figure \ref{fig:ex-r} shows how to obtain $R(T)$ from a simple type graph $T$ using the successive transformations $R_1$, 
$R_2$ and $R_3$.

\begin{figure}[htbp]
\centering
\includegraphics[scale=0.7]{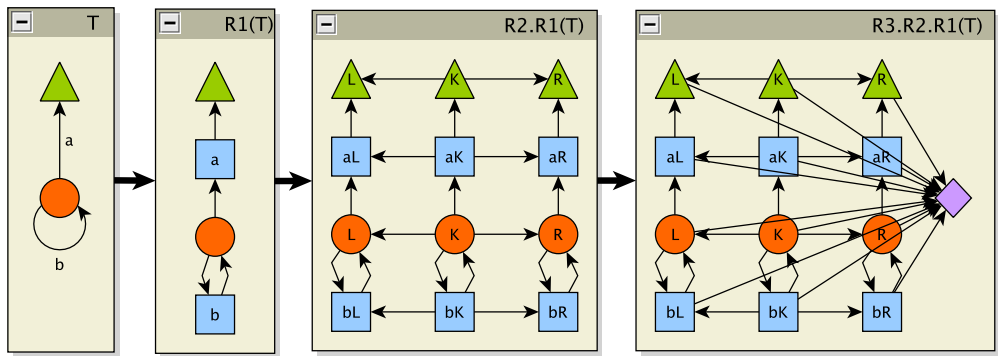}
\caption{Example of type graph rule encoding.}
\label{fig:ex-r}
\end{figure}
\end{example}

Now we define the operation $S$, which encodes $T$-typed rules as $R(T)$-typed graphs. 

\begin{definition}[Graph rule encoding]
Given a $T$-typed graph rule $r: L \stackrel{l}\gets K \stackrel{r}\to R$, the
graph rule encoding $S: Rules(T) \to \Typed{R(T)}$ is defined as the composition
$S_3 \circ S_2 \circ S_1$, where
\bi
\tm $S_1$ applies $R_1$ to $L,K,R$, and types them, respectively, over the left, central and right embeddings of $R_1(T)$ in $R(T)$.
\tm $S_2$ adds concrete edges to the input graph following the homomorphisms $l$ and $r$.
\tm $S_3$ creates a new node typed over the node created by transformation $R_3$, together with new edges connecting all nodes of the input graph to it.
\ei
\end{definition}

\begin{definition}[Graph rule set encoding]
Given a graph grammar $\mc{G} = \< T, G_0, P, \pi\>$, the 
$R(T)$-typed graph $\Upsilon(\mc G)$ is defined as the gluing
$[S(\pi(r_1)),S(\pi(r_2)),\ldots,S(\pi(r_n))]$ where $r_{1..n} \in P$.
\end{definition}

The graph $\Upsilon(\mc G)$ is the union of all encodings of rules $S(\pi(r_x))$ typed
over the single type graph $R(T)$. 

\begin{example}[Graph rule set encoding] 
Figure~\ref{fig:ex-s} depicts a small graph grammar $\mc{G}$ (on the left) and its respective $R(T)$-typed encoding $\Upsilon(\mc G)$ (on the right).

\begin{figure}[htbp]
\centering
\includegraphics[scale=0.7]{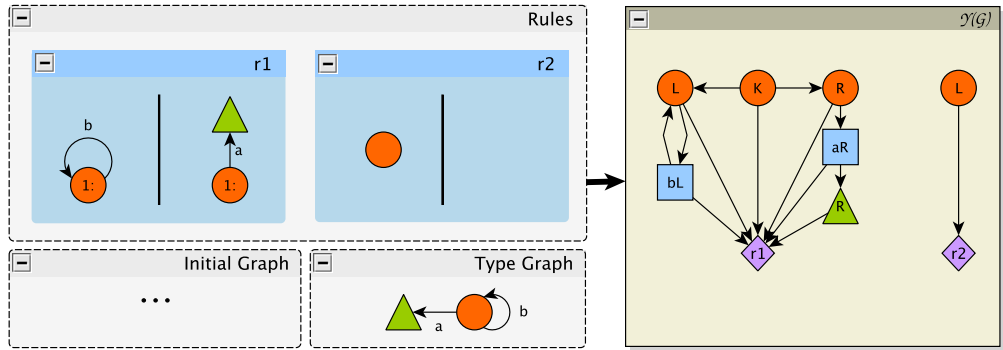}
\caption{Example of graph rule set encoding.}
\label{fig:ex-s}
\end{figure}
\end{example}

\begin{definition}[Encoded graph grammar]
Given a graph grammar $\mc{G} = \< T, G_0, P, \pi\>$, its graph encoding 
is the $(T+R(T))$-typed graph $|\mc{G}| = G_0 + \Upsilon(\mc G)$, where ``$+$'' stands for
the disjoint union of typed graphs. 
\end{definition}

\begin{definition}[Encoded advice]
Let $A = \< D, t, g \>$ be and aspect over $\mc G$, such that $t: T \to T'$. Then, for every
advice $a : p \gets i \to e \in D$, its respective encoding $|a|$ is the $R(T')$-typed span $S(p) \gets S(i) \to S(e)$.
\end{definition}

\begin{definition}[Encoded aspect-oriented graph grammar]
Let $\mc D = (\mc G, \Delta)$ be an AOGG, such that $\mc G = \< T, G_0, P, \pi\>$ and $\Delta = [A_1, A_2, \ldots, A_n ]$.
Let $\mc G^\Delta_W = \< T^\Delta_W, G^\Delta_W, P^\Delta_W, \pi^\Delta_W \>$ be the weaved graph grammar 
obtained from weaving $\Delta$ over $\mc G$. 
Then, the encoded aspect-oriented graph grammar $|\mc D|$ is the graph grammar $\< T^\Delta_W + R(T^\Delta_W), |\mc G'|, |P_{\Delta}|, |\pi_{\Delta}| \>$ where $\mc G' = \langle T^\Delta_W, G^\Delta_W, P, \pi \rangle$ is the type and initial graph extension of $\mc G$ and the rule set $(|P_{\Delta}|,|\pi_{\Delta}|)$ correspond to the gluing of all encoded advices $|a| \in D_i$ of all aspects $A_i = \< D_i, t_i, g_i \>$, $1 \leq i \leq n$. 
\end{definition}

In an AOGG $\mc D$, each aspect extend the type graph and initial graph of the original system. In the encoded graph $|\mc D|$, all these extensions must be present in order to accommodate all encoded advices (of all aspects) in the same rule set.

\begin{figure}[htbp]
\centering
\includegraphics[scale=0.82]{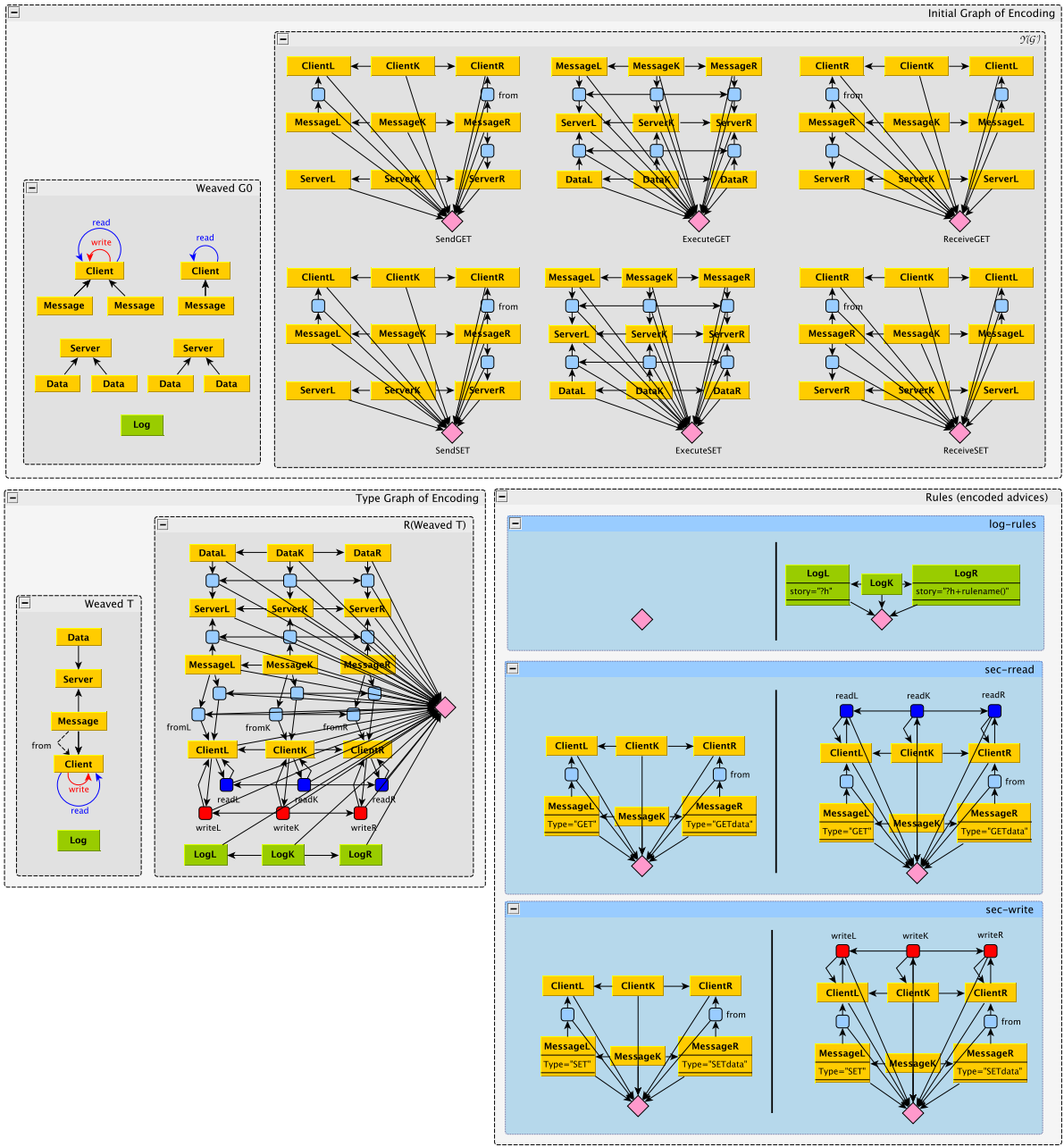}
\caption{Encoded client-server system with log and security aspects.}
\label{fig:encoding}
\end{figure}

\begin{example}[Encoded AOGG]
Figure~\ref{fig:encoding} shows the encoding of $(\mc G,[A_1,A_2])$, where $\mc G$ is the 
base system of Figure~\ref{fig:gragra}, $A_1$ is the log aspect (Figure~\ref{fig:aogg-log})
and $A_2$ is the security aspect (Figure~\ref{fig:aogg-sec}). 
For the sake of readability, some elements are omitted: attribute types in the type graph and attribute values in the initial 
graph and base system rules. However, attribute values are shown in the encoded advices. In all encoded advices the interface graph $K$ and the left-hand side graph $L$ are the same.
\end{example}

\section{Analyzing AOGG specifications}
\label{sec:analyze}

From the point of view of aspect weaving, it is relevant to be able to predict when two different advices
will potentially affect the same part of the specification. If this occurs, 
a different aspect ordering may yield a different weaved system. Given that we encode an AOGG specification
as a graph grammar, we are able to apply techniques from graph rewriting theory to test the existence of conflicts between aspects.
 
Critical pair analysis (CPA) is a static analysis technique that is used to detect both conflicts and dependencies between graph rewriting rules. Given a graph 
$G$ and two direct derivations $G \stackrel{r_1,m_1}{\Longrightarrow} G'$, $G \stackrel{r_2,m_2}{\Longrightarrow} G''$, 
we say that the derivations are in conflict if one of them deletes some graph element that is preserved or deleted by the other one. Thus, their execution in  parallel is not possible. 
Given two subsequent direct derivations $G \stackrel{r_1,m_1}{\Longrightarrow} G' \stackrel{r_2,m_2}{\Longrightarrow} G''$,
we say that there is a dependency between $(r_1,m_1)$ and $(r_2,m_2)$ if the first creates a graph element that is preserved or deleted by the later. In this case, it is not possible to swap their order of execution. Critical pair analysis is used to test potential conflicts and dependencies within a graph grammar and consists of the following general steps: 

\be
 \tm The set of all pairs of rules $(r_1,r_2) \in P \times P$ is generated.
 \tm For each pair $(r_1,r_2)$, all possible overlaps between a graph of $r_1$ and a graph of $r_2$ are calculated. In the case of conflict analysis, both $LHS$s are considered in this calculation. In the case of dependencies analysis, the $RHS$ of $r_1$ and the $LHS$ of $r_2$ are considered. 
 \tm All overlaps are tested for conflicts or dependencies. 
 \tm The number of all conflicting (or dependent) overlaps for each pair of rules is shown in a table.
\ee

\begin{figure}[htb]
\centering
\includegraphics[scale=0.45]{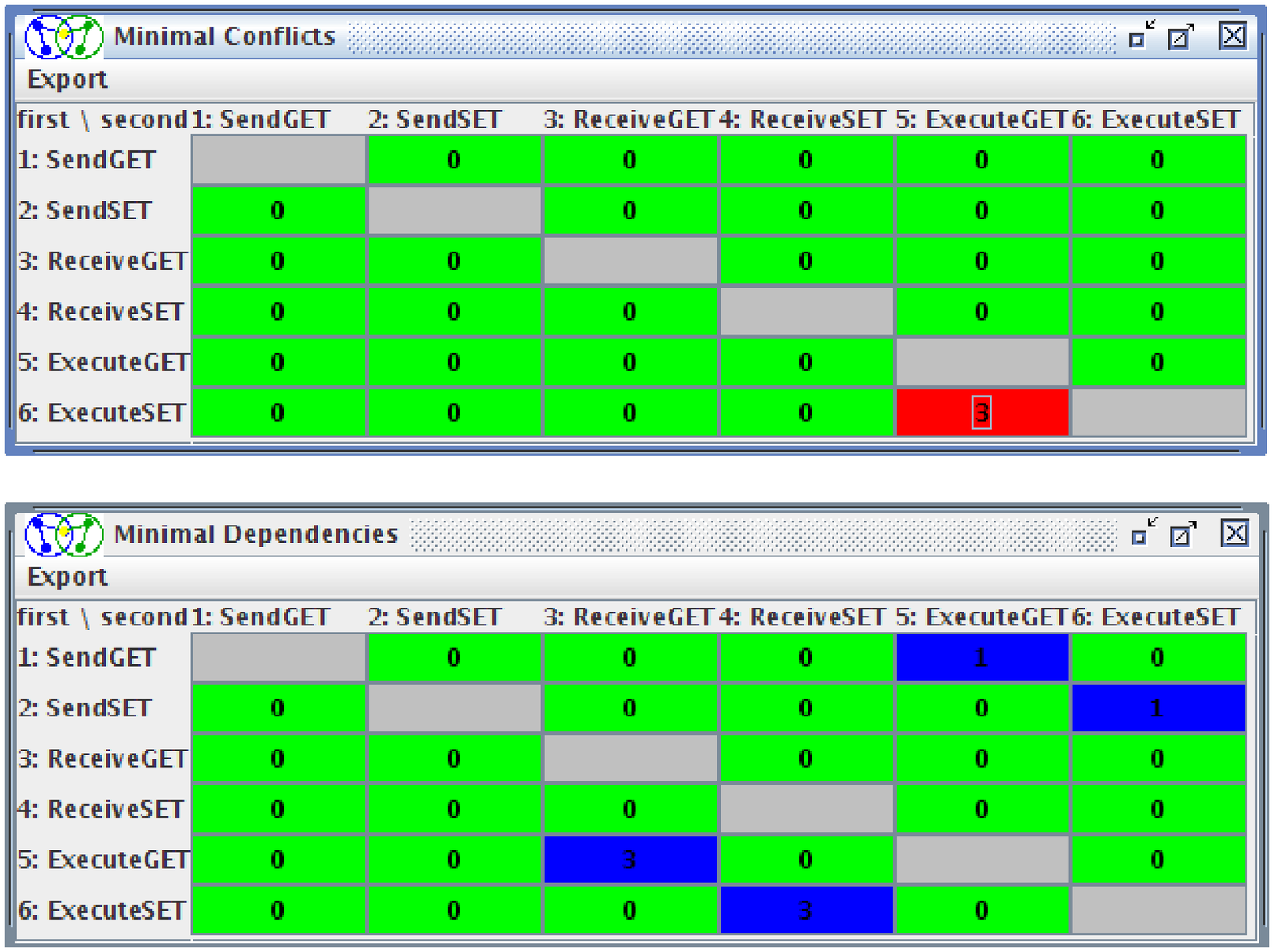}
\caption{Critical pair analysis of the base system in AGG.}
\label{fig:cpa-gg}
\end{figure}

\begin{figure}[htb]
\centering
\includegraphics[scale=0.45]{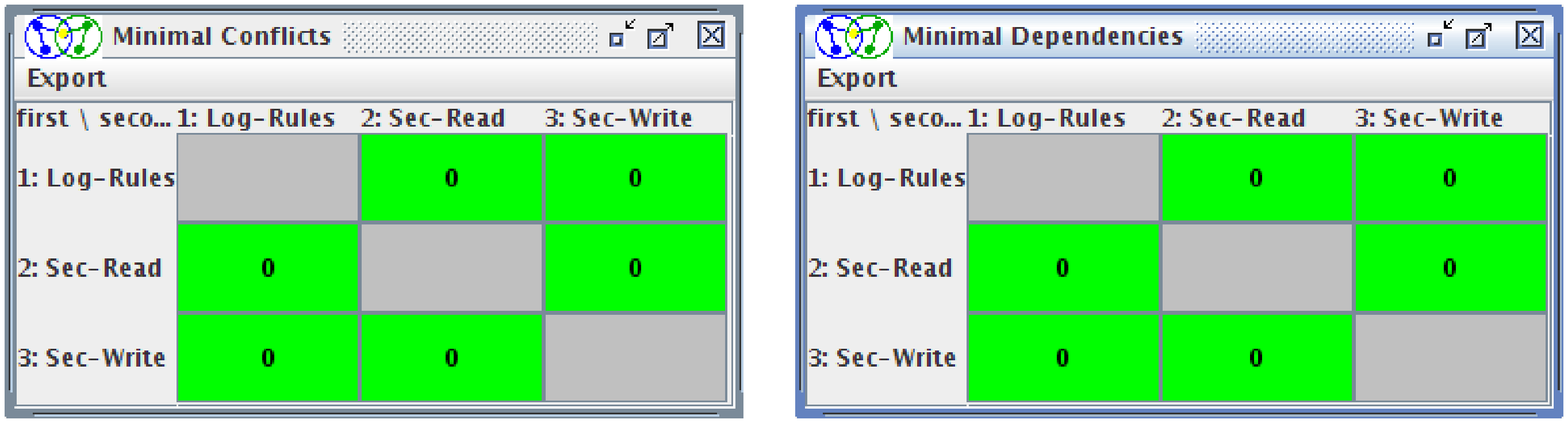}
\caption{Critical pair analysis of the aspect-oriented encoded system in AGG.}
\label{fig:cpa-aogg}
\end{figure}

Aspect-oriented graph grammars are two-layer specifications where the base graph grammar describe the
original system behavior, while aspects describe the implementation of global policies as model transformation. 
By using the encoding of Section 4, we can use AGG to study characteristics of
aspect weaving in a similar way as the original system execution. 
Figure~\ref{fig:cpa-gg} shows the results of critical pair analysis for the base graph grammar.
In this level, conflicts arise only in the pair (\texttt{ExecuteSet},\texttt{ExecuteGET}) when the  first rule updates a value which is read by the second one. Dependencies reflect  \texttt{Send-Execute-Receive} sequences of derivations, induced by the value of the \texttt{Type} attribute in message nodes. In the aspect level we analyze the possibility of
conflicts in the aspect weaving process, since the rules are actually encoded advices. Figure~\ref{fig:cpa-aogg} shows the results of CPA for the encoded model of Figure \ref{fig:encoding}. In this particular system,  where the advices do not delete elements from rules, the rules are obviously conflict-free. Lack of conflicts in the aspect level
means that final weaved system does not change even if we change the order aspects are weaved to the base system. 
If there was at least one conflicting pair of encoded advices $(a_1,a_2)$ where $a_1$ and $a_2$ are not in the same aspect, then the aspects could potentially interact and aspect weaving would not be necessarily a confluent operation.

\section{Final remarks}
\label{sec:conclusion}

This work proposed the use of AOGG to model distributed system with aspects, and also the use of the AGG tool
to study their behavior. We introduced the AOGG model, described an encoding mechanism to describe the base system as a single typed graph, and discussed the use of critical pair analysis to reason about conflicts in the aspect weaving operation. 

The idea of using graph rewriting to define aspect weaving is not entirely new. In \cite{Mehner2006,Whittle2007}, aspect weaving is also defined by means of graph rewriting, and the AGG tool is used to analyze the specification using critical pair analysis. In those approaches, however, UML diagrams are used as base system model. Following a different direction, the work by Aksit et al. \cite{Aksit2009} encodes a complete aspect-oriented calculus into the graph transformation tool Groove \cite{Rensink2004a}. This allows to study the interference between aspects directly in the system unfolding, by employing the space-state generation feature of Groove. 

As future work, we intend to explore the properties of the proposed encoding. This is required in order to relate
the behavior of the weaved system to the behavior of the original base system together with aspect specification.
Furthermore, the usage of graph grammars as base systems may provide insights about how rule-base modification of rules affect the respective rule derivations, since both are representable as diagrams within the same categorical framework.

\ 

\noindent \textbf{Acknowledgements:\ \ }
The authors would like to thank the anonymous referees for their helpful comments and suggestions. This work was partially supported by CNPq – Conselho Nacional de Desenvolvimento Científico e Tecnológico.

\bibliographystyle{eptcs}
\bibliography{database}

\end{document}

%% file: defs.tex
\usepackage[dvipsnames]{xcolor}
\usepackage[all,dvips,crayon]{xy}

\newcommand{\bi}{\begin{itemize}}
\newcommand{\ei}{\end{itemize}}
\newcommand{\be}{\begin{enumerate}}
\newcommand{\ee}{\end{enumerate}}
\newcommand{\bd}{\begin{description}}
\newcommand{\ed}{\end{description}}
\newcommand{\tm}{\item}

\newcommand{\<}{\langle}
\renewcommand{\>}{\rangle}
\newcommand{\mc}[1]{\ensuremath{\mathcal{#1}}}

\newcommand{\Graph}{\ensuremath{\mathbf{Graph}}}

\newcommand{\Typed}[1]{\ensuremath{{#1}\mbox{-}\mathbf{Graph}}}  

\newcommand{\STyped}[1]{\ensuremath{{#1}\mbox{-}\mathbf{MSpan}}}

